\documentclass[letterpaper, 10 pt, conference]{ieeeconf}  

\IEEEoverridecommandlockouts                              

\usepackage{amssymb}
\usepackage{mathrsfs}
\usepackage{algorithm}
\usepackage{algpseudocode}
\usepackage[noadjust]{cite}
\usepackage{bm}
\usepackage{amsmath}
\usepackage{commath}
\usepackage[utf8]{inputenc}
\usepackage{enumerate}
\usepackage{mathtools}

\usepackage{amsthm}
\newtheorem{thrm}{Theorem}

\newtheorem{defn}[thrm]{Definition}

\newtheorem{assumpt}[thrm]{Assumption}

\theoremstyle{definition}
\title{\LARGE \bf
Robust Adaptive Model Predictive Control of Quadrotors
}

\author{Alexandre Didier, Anilkumar Parsi$^{1}$, Jeremy Coulson, Roy S. Smith$^{1}$
\thanks{All authors are with the Department of Information Technology and Electrical Engineering at ETH Zurich, Switzerland {\tt\small\{adidier,aparsi,jcoulson,rsmith\}@ethz.ch.}}
\thanks{$^{1}$Affiliation: This work is supported by the Swiss National Science Foundation under grant number 200021\_178890. The authors are with the Automatic Control Laboratory, ETH Zurich, Switzerland.}%
}

\begin{document}

\maketitle
\thispagestyle{empty}
\pagestyle{empty}

\begin{abstract}
Robust adaptive model predictive control (RAMPC) is a novel control method that combines robustness guarantees with respect to unknown parameters and bounded disturbances into a model predictive control scheme. 
However, RAMPC has so far only been developed in theory. The goal of this paper is to apply RAMPC to a physical quadrotor experiment. To the best of our knowledge this is the first time that RAMPC has been applied in practice using a state space formulation. In doing so, we highlight important practical challenges such as computation of $\lambda$-contractive polytopes and dealing with measurement noise, and propose modifications to RAMPC so that it can be applied on a quadrotor.
We first simulate quadrotor flight with a direct and a decoupled control architecture in different scenarios. The scenarios include: (i) an uncertain quadrotor mass and additive wind disturbance as part of a package delivery problem; and (ii) all rotor efficiencies drop as a power delivery problem. 
We then implement these scenarios on a physical quadrotor and present the experimental results.
\end{abstract}

\section{INTRODUCTION}
\label{sec:intro}
Model Predictive Control (MPC), see e.g. \cite{kouvaritakis2016model}, is an optimisation based control scheme, which guarantees state and input constraint satisfaction for a discrete-time, nominal system. Robust MPC, see \cite{kouvaritakis2016model}, integrates robustness guarantees with respect to bounded disturbances and model uncertainties into the optimisation problem. However, robust MPC controllers result in poor closed loop performance when the uncertainty is large. This is addressed by Robust Adaptive MPC (RAMPC), a novel control technique proposed in \cite{tanaskovic2014adaptive} using impulse response models to describe system dynamics. The method assumes that the impulse response coefficients lie in a bounded set describing the model uncertainty. Using measurement data, this uncertainty set is updated online by applying set-membership identification \cite{milanese1991optimal}. By ensuring that the true parameter is always included in the updated uncertainty set, robust constraint satisfaction is guaranteed in closed loop. The RAMPC method was extended in \cite{lorenzen2019robust} for linear state space models with affine parametric uncertainty in the state space matrices. This extension enables the use of fewer parameters to describe the model uncertainty, and allows to apply RAMPC to a wider class of systems. Multiple extensions have been proposed to this algorithm to reduce its computational complexity, see e.g. \cite{kohler2019linear,lu2019robust}. Despite these attractive features, RAMPC has never been applied in practice using state space models. In this paper, we aim to bridge the gap between theory and practice, by implementing RAMPC on a quadrotor to perform regulation in various scenarios. The algorithm robustly guarantees a safe flight when the model parameters are not accurately known and the system is affected by disturbances. This provides an inherent safety advantage compared to other adaptive schemes which have been applied to quadrotors, see e.g. \cite{chikasha2017adaptive}. 


%

The main contributions of this paper are the solutions to practical issues which arise in the application of RAMPC. The first issue is that the steady-state input is itself not known, which is generally ignored when a problem is formulated for regulation. The second issue is that the RAMPC algorithm for state space models assumes that the measurements are unaffected by noise, which is not realistic. Finally, the algorithm also requires computation of contractive polytopes, whose design affects the feasible region and closed loop performance. To resolve these issues, we present solutions with theoretical guarantees which showed good performance during practical implementation. In addition, we propose a novel algorithm to compute polytopes with a desired contraction rate, which reduces conservatism and computation time of the algorithm.

To test the effectiveness of RAMPC as a control scheme for quadrotors, we first simulate different scenarios of a quadrotor flight: (i) The first scenario involves an uncertain quadrotor mass as part of a package delivery. If the mass of the package is unknown, it may lead to an unsafe flight resulting in a crash of the quadrotor. Additionally, an unknown wind force is acting on the quadrotor which is modelled as an additive disturbance. Wind is a common factor during an outdoor quadrotor flight and its impact on quadrotor flight is studied in \cite{zhang2016three} and \cite{waslander2009wind}. (ii) In the second scenario we consider a loss of efficiency in all rotors as a power delivery problem, as similarly studied in \cite{amoozgar2013experimental} and \cite{dydek2010combined}. The simulations were performed with two different control architectures for the quadrotor, a direct thrust control and a decoupled control structure, see \cite{nrotorscript}.

The paper is structured as follows. Section \ref{sec:RAMPC} contains a description of the RAMPC method used, a description of the set-membership identification, and a discussion on practical implementation issues and their solutions. In Section \ref{sec:simulations}, we show the results of simulations of different quadrotor flight scenarios and in Section \ref{sec:experiments} we discuss the results of the experiments on the physical quadrotor.

\textit{Notation.} The set of integers $\{1,\dots,n\}$ is given as $\mathbb{N}_1^n$ and the set of positive reals is $\mathbb{R}_{>0}$. $A\succeq0$ denotes that the matrix $A$ is positive semi-definite. The $i$-th row of a matrix $A$ is given as $[A]_i$. The diagonal matrix with entries $a,b,c$ on its main diagonal is denoted as $\textup{diag}(a,b,c)$ and the convex hull of a set $\mathbb{A}$ is defined as $\text{co}(\mathbb{A})$. The identity matrix is denoted as $\mathbb{I}$ and $\mathbf{1}$ is the vector of ones. The Euclidean norm of $b$ is is given by $\norm{b}$ and $\norm{x}_A^2$ represents $x^TAx$. $A\oplus B$ denotes Minkowski set addition and $\mathbb{B}_n$ denotes the set $\{x\in\mathbb{R}^n|\;\norm{x}_\infty\leq0.5\}$.

\section{Robust Adaptive Model Predictive Control}\label{sec:RAMPC}
\subsection{System description}
The RAMPC scheme proposed in \cite{kohler2019linear} is used in this paper due to its computational efficiency. The method considers linear, discrete-time, parameter dependent dynamics with an additive disturbance
\begin{equation}\label{eq:dyn}
x_{k+1}=A(\theta)x_k+B(\theta)u_k+w_k,
\end{equation}
with the state vector $x_k\in\mathbb{R}^n$, the disturbance $w_k\in\mathbb{R}^n$, the input vector $u_k\in\mathbb{R}^m$ and the uncertain parameter $\theta\in\mathbb{R}^p$, whose true value is $\theta=\theta^*$. The additive disturbance $w_k$ and the uncertain parameter $\theta$ fulfil the following assumptions, made in \cite{lorenzen2019robust}, \cite{lu2019robust} and \cite{kohler2019linear}:
\begin{assumpt}\label{ass:bounded} \textup{ }
\begin{enumerate}
\item The disturbance $w_k$ is bounded by a convex polytope 
\begin{equation}
w_k\in\mathbb{W}=\{w\in\mathbb{R}^n |\; H_ww\leq h_w\},
\end{equation}
with $H_w\in\mathbb{R}^{n_w\times n}$ and $h_w\in\mathbb{R}^{n_w}$. 
\item The system matrices $A(\theta)$ and $B(\theta)$ depend affinely on the parameter vector $\theta\in\mathbb{R}^p$ with respect to the matrices $A_i\in\mathbb{R}^{n\times n}$ and $B_i\in\mathbb{R}^{n\times m}$, such that
\begin{equation}
(A(\theta),B(\theta))=(A_0,B_0)+\sum_{i=1}^p(A_i,B_i)[\theta]_i.
\end{equation}
\item The parameter $\theta$ is bounded in a convex polytope 
\begin{equation}
\theta\in\Theta_0=\{\theta\in\mathbb{R}^p |\; H_{\theta_0}\theta\leq h_{\theta_{0}}\},
\end{equation}
with $H_{\theta_0}\in\mathbb{R}^{n_\theta \times p}$ and $h_{\theta_0}\in\mathbb{R}^{\theta}$ known, containing the true, unknown parameter vector $\theta^*$.
\end{enumerate}
\end{assumpt}

The states and inputs are constrained in a bounded polytopic set, with $F\in\mathbb{R}^{n_z\times n}$ and $G\in\mathbb{R}^{n_z\times m}$,
\begin{equation}\label{eq:stateinputconstr}
(x_k,u_k)\in\mathbb{Z}=\{(x,u)\in\mathbb{R}^n\times\mathbb{R}^m|\;Fx+Gu\leq \mathbf{1}\}.
\end{equation}

\subsection{Parameter Identification}\label{sec:paramid}
To estimate the parameter $\theta$, set-membership identification is used, see \cite{milanese1991optimal}. 
In this identification method, a set of possible parameters that is guaranteed to contain the true parameter is recursively updated. This gives robustness guarantees with respect to the uncertain parameter. In order to compute this recursive update, the time dependent matrix $D_k\in\mathbb{R}^{n\times p}$ and vector $d_k\in\mathbb{R}^n$ are introduced
\begin{gather}
\begin{align}
D_k&=[A_1x_k{+}B_1u_k, A_2x_k{+}B_2u_k, \dots , A_px_k{+}B_pu_k], \nonumber\\
d_k&=A_0x_{k-1}+B_0u_{k-1}-x_k, \nonumber
\end{align}
\end{gather}
which are affine in the states and inputs.
At every time step $k$, the set of possible parameters consistent with the evolution of the system given the unknown, bounded disturbance, can be computed. This set of parameters $\Delta_k$, called the non-falsified parameter set, is given as
\begin{gather}
\begin{align}
\Delta_k & =\{\theta\in\mathbb{R}^p |\; x_k-(A(\theta)x_{k-1}+B(\theta)u_{k-1})\in\mathbb{W}\} \nonumber \\
&=\{\theta\in\mathbb{R}^p |\; -H_wD_{k-1}\theta\leq h_w+H_wd_k\}. 
\end{align}
\end{gather}
The set of feasible parameters is denoted by $\Theta_k$ and is recursively updated using the non-falsified set $\Delta_k$, starting with $\Theta_0$ as the initial set. In order to efficiently describe the parameter set, the parameter sets are restricted to a bounding hypercube, with centre $\bar{\theta}_k\in\mathbb{R}^p$ and length $\eta_k\in\mathbb{R}_{\geq0}$, as in
\begin{equation}
\Theta_k=\{\bar{\theta}_k\}\oplus\eta_k\mathbb{B}_p\supseteq\Theta_{k-1}\cap\Delta_k .
\end{equation}
The computation of such $\bar{\theta}_k$ and $\eta_k$ is detailed in \cite{kohler2019linear}.
Additionally, a point estimate $\hat{\theta}_k$ is required to evaluate the cost function, which is chosen as a least mean squares estimate of the true parameter $\theta^*$ and is updated recursively using the update formula
\begin{gather}
\tilde{\theta}_k=\hat{\theta}_{k-1}{+}\mu D_{k-1}^T(x_k{-}(A(\hat{\theta}_{k-1})x_{k-1}{+}B(\hat{\theta}_{k-1})u_{k-1})), \nonumber\label{eq:thetahatupdate} \\
 \hat{\theta}_k=\Pi_{\Theta_k}(\tilde{\theta}_k), \label{eq:thetaproj}
\end{gather}
where $\Pi_{\Theta_k}(\cdot)$ denotes the projection onto the set $\Theta_k$ and $\mu\in\mathbb{R}_{>0}$ is a constant filter parameter. 

\subsection{Tube Model Predictive Control}
Tube model predictive control, see \cite{langson2004robust}, predicts the state propagation of a system by using polytopes within which the states are guaranteed to be contained, given a bounded additive disturbance and bounded uncertain parameters. In order to guarantee robustness with respect to $w\in \mathbb{W}$ and $\theta\in \Theta_k$, a state tube $\{\mathbb{X}_{l|k}\}_{l\in\mathbb{N}_0^N}$ is introduced at time step $k$ for the predicted time step $l$. This tube consists of $N+1$ predicted polytopes $\mathbb{X}_{l|k}\subset\mathbb{R}^n$ in the state space at time step $k$, where $N$ is the prediction horizon of the RAMPC scheme. At each predicted time step $l|k$, it can be ensured that the states are inside of the polytope $\mathbb{X}_{l|k}$ for all $w\in\mathbb{W}$ and $\theta\in\Theta_k$. It follows that robustness is guaranteed if the polytopes do not violate the constraints in \eqref{eq:stateinputconstr}. Thus, the following constraints must hold $\forall l\in\mathbb{N}_0^N$
\begin{subequations} \label{eq:constraints}
\begin{alignat}{1}
& A(\theta)x+B(\theta)u_{l|k}(x)+w \in \mathbb{X}_{l+1|k}, \nonumber\\
&\qquad\qquad \forall x\in\mathbb{X}_{l|k}, w\in\mathbb{W}, \theta\in\Theta_k,\label{eq:tubeinc} \\
&x_k \in\mathbb{X}_{0|k}, \quad (x,u_{l|k}(x))\in\mathbb{Z}, \quad \forall x \in\mathbb{X}_{l|k}, \label{eq:stateconstr}
\end{alignat}
\end{subequations}
for some input mapping $u_{l|k}: \mathbb{R}^n\rightarrow \mathbb{R}^m$. 
The polytopes $\mathbb{X}_{l|k}=\{\bar{x}_{l|k}\}\oplus\alpha_{l|k}\mathbb{X}_0$ are defined as translations and dilations of a predefined polytope $\mathbb{X}_0=\{x\in\mathbb{R}^n|\;H_xx\leq\mathbf{1}\}$, with $H_x\in\mathbb{R}^{n_x\times n}$. The translations $\bar{x}_{l|k}$ are computed according to the dynamics \eqref{eq:dyn} with $\theta=\bar{\theta}_k$ and the dilations $\alpha_{l|k}$ are decision variables.
Finally, in order to guarantee recursive feasibility of the scheme, the following terminal constraint is applied on the final polytope of the tube $\mathbb{X}_{N|k}\subseteq \mathbb{X}_f$, where $\mathbb{X}_f$ is a terminal set.

An input parametrisation $u_{l|k}(x)=Kx+v_{l|k}$ is used, with a prestabilising feedback matrix $K\in\mathbb{R}^{m\times n}$ and the decision variables $\{v_{l|k}\}_{l\in\mathbb{N}_0^{N-1}}$, where $v_{l|k}\in\mathbb{R}^m$. The feedback matrix $K$ needs to fulfil the following assumption. 

\begin{assumpt} \label{ass:prestab} 
The feedback gain K stabilises $A_{cl}(\theta){=}A(\theta){+}B(\theta)K$ for all $\theta{\in}\Theta_0$ and it holds that  
\begin{equation}\label{eq:prestab}
A_{cl}(\theta)^TPA_{cl}(\theta)+Q+K^TRK\succeq P, \quad\forall \theta\in\Theta_0,
\end{equation} 
where $Q\in\mathbb{R}^{n\times n},R\in\mathbb{R}^{m\times m}$ and $P\in\mathbb{R}^{n\times n}$ are positive definite cost matrices of the cost function defined in \eqref{eq:optcost}. 
\end{assumpt} 

Assumption 2 is standard for tube MPC methods, and is also used in \cite{lorenzen2019robust} and \cite{kohler2019linear}. The prestabilizing gain K and the terminal cost $P$ can be computed using a semi-definite program, for example as proposed in \cite{kohler2019linear}.

The RAMPC method in \cite{kohler2019linear} uses the contractivity of the polytope $\mathbb{X}_0$ 
in order to rewrite the constraints \eqref{eq:constraints} linearly in the optimisation variables. The definition of a $\lambda$-contractive polytope is given as follows.
\begin{defn}\label{def:lambda}
A polytopic set $\mathbb{X}_0=\{x\in\mathbb{R}^n | \;H_xx\leq \mathbf{1}\}$ is $\lambda$-contractive for some $\lambda\in[0, 1)$, with respect to some $\theta\in\Theta_0$ and a feedback gain $K$, if
\begin{equation*}
H_x(A(\theta)+B(\theta)K)x\leq\lambda\mathbf{1},\quad \forall x\in\mathbb{X}_0.
\end{equation*}
\end{defn}

In order to compute the contractivity of a polytope $\mathbb{X}_0$ for all $\theta\in\Theta_k\subseteq\Theta_0$, the upper bound, denoted $\bar{\lambda}$,
\begin{gather*}
\begin{align}
\bar{\lambda}{=}\max_{i,x\in\mathbb{X}_0}[H_x]_iA_{cl}(\bar{\theta}_k)x{+}\eta_k\max_{i,j,x\in\mathbb{X}_0}[H_x]_iD(x,Kx)\tilde{e}_j, \nonumber 
\end{align}
\end{gather*}
is used, where $\tilde{e}_j$ represents the $j$-th vertex of the unit hypercube $\mathbb{B}_p$.  
The RAMPC optimisation problem, which is solved at every time step, using $\alpha_{\cdot|k}{=}\{\alpha_{l|k}\}_{l=0,\dots,N}$ and $v_{\cdot|k}{=}\{v_{l|k}\}_{l=0,\dots,N-1}$ is:
\begin{subequations} \label{eq:efftubeopt}
\begin{align}
\min_{v_{\cdot|k}, \alpha_{\cdot|k}} & \sum_{l=0}^{N-1}\norm{\hat{x}_{l|k}}^2_Q+\norm{\hat{u}_{l|k}}_R^2+\norm{\hat{x}_{N|k}}^2_P \label{eq:optcost}\\
\textup{s.t. }& \forall i\in\mathbb{N}_1^{n_x},\forall j\in\mathbb{N}_1^{2^p},l \in\mathbb{N}_0^{N-1}, \nonumber\\
& \bar{x}_{0|k}=\tilde{x}_{0|k}=x_k, \alpha_{0|k}=0, \label{eq:optinit}\\
& \bar{x}_{l+1|k}=A(\bar{\theta}_k)\bar{x}_{l|k}+B(\bar{\theta}_k)\bar{u}_{l|k},\label{eq:optbar1}\\
& \bar{u}_{l|k}=K\bar{x}_{l|k}+v_{l|k},  \label{eq:optbar2}\\
& \hat{x}_{l+1|k}=A(\hat{\theta}_k)\hat{x}_{l|k}+B(\hat{\theta}_k)\hat{u}_{l|k}, \label{eq:opthat1}\\
& \hat{u}_{l|k}=K\hat{x}_{l|k}+v_{l|k},  \label{eq:opthat2}\\
& (F+GK)\bar{x}_{l|k}+Gv_{l|k}+c\alpha_{l|k}\leq\mathbf{1},  \label{eq:optconstr} \\
& \bar{\lambda}\alpha_{l|k}{+}\bar{w}{+}\eta_k[H_x]_iD(\bar{x}_{l|k},\bar{u}_{l|k})\tilde{e}_j\leq\alpha_{l+1|k}, \label{eq:opttubeinc}\\
& (\alpha_{N|k}+[H_x]_i\bar{x}_{N|k})\max_i[c]_i\leq1 \label{eq:optterm}
\end{align}
\end{subequations}
with $\bar{w}=\max_{i,w\in\mathbb{W}}[H_x]_iw$ and $[c]_i=\max_{x\in\mathbb{X}_0}[F+GK]_ix$ for $i\in\mathbb{N}_1^{n_z}$. The optimisation problem consists of the cost function \eqref{eq:optcost}, the initial condition \eqref{eq:optinit}, the propagation of the state $\bar{x}$ for the centre $\bar{\theta}_k$ of $\Theta_k$ in \eqref{eq:optbar1} and \eqref{eq:optbar2} and of $\hat{x}$ for the point estimate $\hat{\theta}$ in \eqref{eq:opthat1} and \eqref{eq:opthat2}, the state and input constraints \eqref{eq:optconstr}, the tube inclusion constraint \eqref{eq:opttubeinc} which implies \eqref{eq:tubeinc} and finally the terminal constraint \eqref{eq:optterm}. Under the assumption that $\bar{\lambda}+\max_ic_i\bar{w}\leq1$ holds, it is shown in \cite{kohler2019linear} that \eqref{eq:optterm} can be used as a terminal constraint to show recursive feasibility. Thus, with the given assumptions the constraints \eqref{eq:constraints} hold and recursive feasibility, stability and consistency of the parameter estimation are proven in \cite{kohler2019linear}. 

\subsection{Practical Issues}\label{sec:pracproblems}
Steady-state input error and measurement noise affect many real systems. For the quadrotor simulations and implementations in Sections \ref{sec:simulations} and \ref{sec:experiments}, an uncertain mass is used, which results in a steady-state input dependent on the uncertain parameter $\theta$ for a linearisation of the dynamics around the hover position. This practical issue is not considered in the methods presented in \cite{lorenzen2019robust}, \cite{lu2019robust} and \cite{kohler2019linear}. Given a system with a true, non-zero steady state input $u_{ss}(\theta^*)$ the steady-state input which is applied is $u_{ss}(\theta)=u_{ss}(\theta^*)+u_{ss,err}(\theta)$ with the steady-state input error $u_{ss,err}(\theta)$. 

This steady-state input error affects the dynamics given by $x_{k+1}=A_{cl}(\theta^*)x_{k} +B(\theta^*)v_{k} +B(\theta^*)u_{ss,err}(\theta) + w_k$.
In order to be able to guarantee robustness with respect to this steady-state input error, the term $B(\theta^*)u_{ss,err}(\theta)$ can be considered as an additional disturbance that affects the dynamics. By using $[\tilde{u}]_i=\max_{\theta\in\Theta_0}[H_x]_iB(\theta)u_{ss,err}(\theta)$ as the rows $i$ of the vector $\tilde{u}$, the tube inclusion constraint can be rewritten for $ x_{l|k}\in\mathbb{X}_{l|k}, \theta\in\Theta_k$ as
\begin{gather}
\begin{align}
H_x\left((A(\theta)+B(\theta)K) x_{l|k} +B(\theta)v_{l|k}\right) + \tilde{w} + \tilde{u}\leq \alpha_{l+1|k}, \nonumber
\end{align}
\end{gather}
 By including this additional term $\tilde{u}$, the theoretical robustness guarantees of the method are preserved. In practice however, this leads to conservatism, as a large uncertainty set $\Theta_0$ is used for the scenarios, which in turn leads to a large steady-state error compared to the disturbance. Instead, as the estimate of $\theta$ improves over time by using the set-membership identification, the steady-state input is recomputed as $u_{ss}(\bar{\theta}_k)$, with $\bar{\theta}_k$ the centre of $\Theta_k$. This significantly improves performance at the cost of the loss of robustness guarantees. Compared to directly using the methods in \cite{lorenzen2019robust}, \cite{lu2019robust} and \cite{kohler2019linear}, updating the steady-state input with respect to $\bar{\theta}_k$ resulted in reference tracking with small or no steady-state error, as can be seen in Sections \ref{sec:simulations} and \ref{sec:experiments}.
 
The second practical issue to be considered is measurement noise on the state $\tilde{x}_k=x_{k}+m_k$. Such noisy measurements can lead to the true parameter $\theta^*$ being removed from $\Theta_k$ in the set-membership identification. Under the assumption of bounded noise $m_k\in\mathbb{M}\subset\mathbb{R}^n$, with a convex polytope $\mathbb{M}$, consistency for the set-membership identification can still be guaranteed. It must hold that if there exists a noise $m\in\mathbb{M}$, that could explain a parameter choice $\theta\in\Theta_{k-1}$ given the disturbance $w\in\mathbb{W}$, then this parameter $\theta$ cannot be eliminated from $\Theta_k$. This can be achieved by introducing dilation factors in the calculation of the non-falsified parameter set
\begin{gather}\label{eq:SMnoise}
\begin{align}
\tilde{\Delta}_k=\{\theta\in\mathbb{R}^n |\; &{-}H_wD_{k-1}\theta\leq h_w{+}H_wd_k{+}\max_{m_k\in\mathbb{M}}H_wm_k \nonumber \\
&+\max_{\theta\in\Theta_{k-1}, m_{k-1}\in\mathbb{M}}-H_wA(\theta)m_{k-1}\},
\end{align}
\end{gather}
with the set update $\Theta_k\supseteq\Theta_{k-1}\cap\tilde{\Delta}_k$. As $\mathbb{M}$ is known a priori and and the sets $\Theta_k$ are restricted to hypercubes, the dilation factors can be easily computed. By using the proposed methods, RAMPC can be applied to any practical system subject to an uncertain steady-state input and bounded measurement noise.

In order to reduce the computation time in the RAMPC scheme, the upper bound $\bar{\lambda}$ in \eqref{eq:opttubeinc} is not updated during real-time control in Section \ref{sec:experiments}. As this results in a conservative upper bound on the contractivity rate, 
the algorithm from \cite{pluymers2005efficient} is adapted to construct a polytope $\mathbb{X}_0$ with a desired contractivity rate $\lambda$ for all $\theta\in\Theta_0$, and is given in Algorithm 1. The specified contractivity rate $\lambda$ replaces the potentially conservative upper bound $\bar{\lambda}$ and reduces the initial conservatism on the contractivity, which results in better initial feasibility. Additionally, by using the proposed algorithm, $\lambda$-contractive polytopes with a low number of half-spaces were constructed, which allowed a real-time application of RAMPC in Sections \ref{sec:simulations} and \ref{sec:experiments}.

\begin{algorithm}[h]
\caption{Computation of a $\lambda$-contractive polytope.}\label{alg:1}
\begin{algorithmic}[1]
\State Initialise $H_x^0=\begin{bmatrix}F^T & (GK)^T\end{bmatrix}^T$ 
\State $i\leftarrow1$
\While{$i\leq$ \# rows of $H_x^{i-1}$}
\State $H_x^i\leftarrow H_x^{i-1}$
\For{$j\in\mathbb{N}_1^{n_{v,\theta}}$}
\State $e_j\leftarrow\max_x [H_x^{i}]_iA_{cl}(\theta^j)-\lambda1$ 
\State \qquad\quad \textup{s.t. } $H_x^{i}x\leq\mathbf{1}$
\If{$e_j>0$}
\State $H_x^{i}=\begin{bmatrix} H_x^{iT} & (\frac{1}{\lambda}[H_x^i]_iA_{cl}(\theta^j))^T \end{bmatrix}^T$
\EndIf
\EndFor
\State $i\leftarrow i+1$
\EndWhile
\State $\mathbb{X}_0=\{x\in\mathbb{R}^n |\; H_x^{i-1}x\leq\mathbf{1}\}$
\end{algorithmic}
\end{algorithm}

\section{Simulation Studies} \label{sec:simulations}
\subsection{Quadrotor Dynamics}\label{sec:quaddyn}
The dynamics of a quadrotor are nonlinear and are represented by 12 states and 4 inputs and their description can be found in \cite{nrotorscript}. The states and inputs are 
\begin{align*}
x=\begin{bmatrix} \Delta p^T & \Delta\dot{p}^T & \Delta \psi^T & \Delta\dot{\psi}^T \end{bmatrix}^T, \\
 u=\begin{bmatrix} \Delta f_1^T & \Delta f_2^T & \Delta f_3^T & \Delta f_4^T \end{bmatrix}^T,
\end{align*}
where $\Delta p$ is the $x$-,$y$-,$z$-positional deviation from the steady-state position, $\Delta \psi = \begin{bmatrix}\Delta\gamma &\Delta\beta & \Delta\alpha\end{bmatrix}^T$ are the roll-pitch-yaw angles and $\Delta f_i$ are the deviations of the thrusts generated by rotor $i$ from a steady-state input. 
For the RAMPC scheme described in Section \ref{sec:RAMPC}, the linearisation around the hover position of the quadrotor dynamics is used.
This steady-state input for the quadrotor is computed by solving 
\begin{equation}\label{eq:ss}
\begin{bmatrix} 1 & 1& 1& 1 \\ y_1 & y_2 & y_3 & y_4 \\ -x_1 & -x_2 & -x_3 & -x_4\\c_1 & c_2 & c_3 & c_4 \end{bmatrix}\begin{bmatrix} f_1 \\ f_2 \\ f_3 \\f_4 \end{bmatrix} = \begin{bmatrix} mg \\ 0 \\0\\0 \end{bmatrix},
\end{equation}
with the position of rotor $i$ with respect to the centre of gravity $(x_i,y_i)$, a constant of proportionality from rotor torque to thrust force $c_i$, the quadrotor mass $m$ and the gravitational acceleration $g$. The linearised dynamics are in the form \eqref{eq:dyn} and are given in \cite{nrotorscript}. For the discrete time dynamics, an Euler discretisation with a sample time of $T_s=0.1\textup{s}$ is used. The chosen discretisation and sampling time showed good flight performance in simulation.

\subsection{Unknown Mass Scenario} \label{sec:unknownmass}
The first scenario considered is of a package delivery, where the mass of the quadrotor $m$ is unknown. For this, we use RAMPC in a receding horizon fashion for a control architecture where the control inputs are the individual thrusts of each rotor as described in Section \ref{sec:quaddyn}, referred to as direct thrust control in \cite{nrotorscript}. Note that as $m$ is required in the computation of the steady-state input \eqref{eq:ss}, $u_{ss}(\theta)$ is updated at every time step as discussed in Section \ref{sec:pracproblems}. The inverse of the mass appears in the dynamics. Thus, $\theta=\frac{1}{m}$ and $\Theta_0=[\frac{1}{0.037},\frac{1}{0.027}]\textup{kg}^{-1}$ with $\theta^*=\frac{1}{0.028}$. 
A constant wind disturbance with a velocity of up to $2\frac{m}{s}$ in $x,y$ and $z$-direction is considered. The quadrotor is restricted to operate in a hypercube in space of $\{\Delta p\in\mathbb{R}^3 |\; \norm{\Delta p}_\infty\leq 0.7\textup{m}\}$. The roll, pitch and yaw angles are restricted to $\pm\pi/2$ radians. As input constraints, the generated rotor thrust for each rotor needs to lie within $[0,0.16]\textup{N}$. The cost matrices, which are used are
Q=\textup{diag}(10,10,100,1,1,1,2,2,30,1,1,1)/100 and
R=\textup{diag}(1,1,1,1)/100.
All optimisation problems are solved using YALMIP \cite{lofberg2004yalmip} with MOSEK \cite{andersen2000mosek} and OSQP \cite{stellato2020osqp} as solvers. The results of this scenario can be seen in Figure 1, where after 10 time steps, $\Theta_{10}\approx[35.66, 37.04]\textup{kg}^{-1}$, which corresponds to $m\in[27,28.05]\textup{g}$. Note that by using the direct thrust control mode, robust flight can be ensured in the lateral directions as well as the altitude. The average solve time for solving the optimisation problem as well as updating the uncertain parameter was $90$ms on a 3.1 GHz Intel i5 CPU. The computation time of $90$ms was achieved by using Algorithm 1 in order to find a $\lambda$-contractive polytope $\mathbb{X}_0\subset\mathbb{R}^{12}$ with a low number of half-spaces $n_x$, as the number of constraints in \eqref{eq:efftubeopt} is dependent on $n_x$.
In Figure 2, the performance of RAMPC is compared to a robust MPC controller while performing the package delivery task. Note that the mass of the quadrotor is assumed to be $37\textup{g}$ for the robust MPC scheme, which lies in $\Theta_0$ and is not updated during the flight. As opposed to the results shown in Figure 1, the robust MPC controller is tracking only the altitude reference as tracking the $x$- and $y$-position at the same time resulted in unstable flight. The robust MPC problem is infeasible near the constraints and a steady-state error exists near the origin, as the steady-state input is not updated as the mass is not estimated.
\begin{figure}[h]
    \centering
    \includegraphics[width=0.48\textwidth]{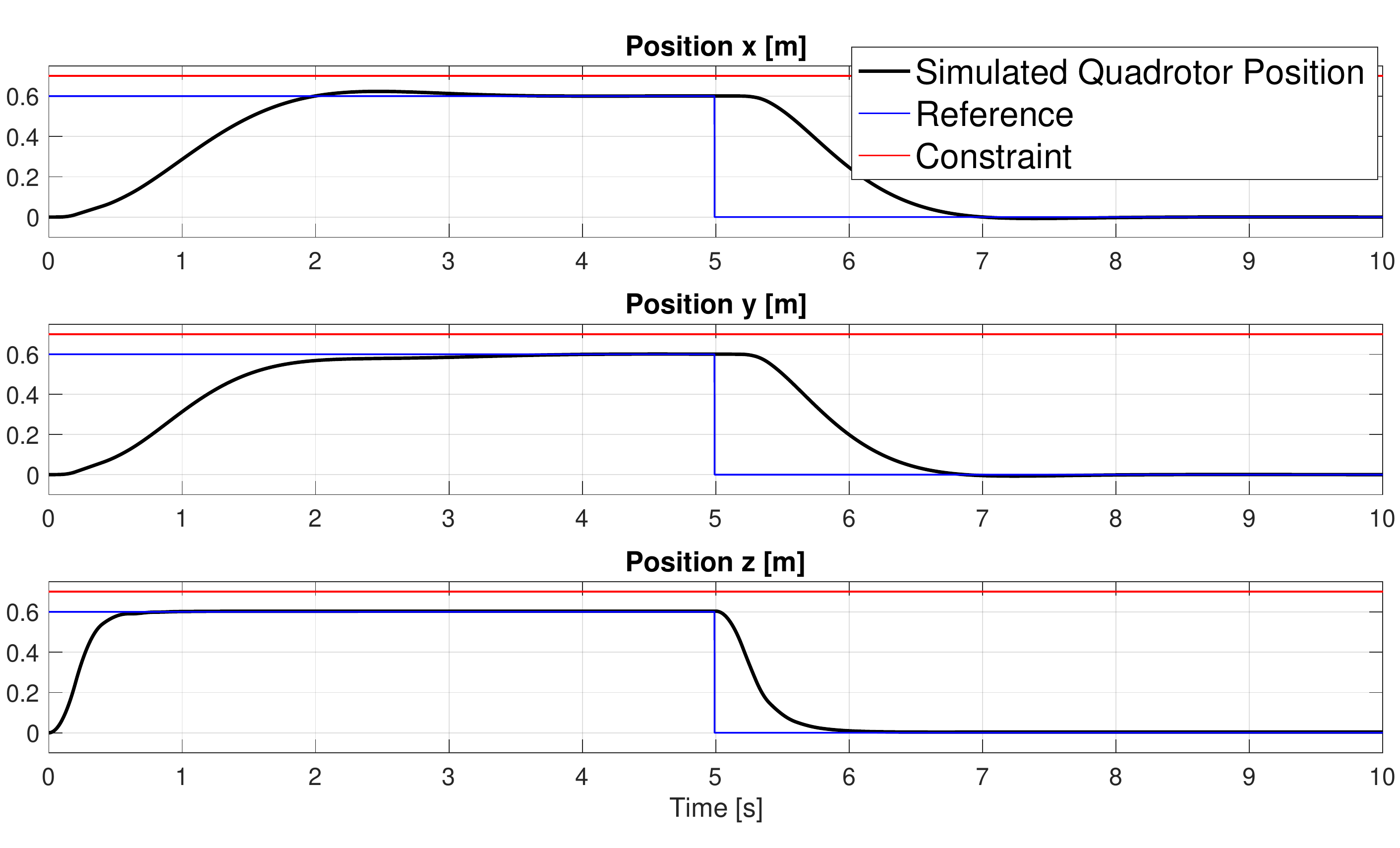}
    \vspace{-0.6cm}
    \caption{Simulation of RAMPC applied to a quadrotor with an uncertain mass and constant wind as a disturbance with position states, constraints and reference.}
\end{figure}
\begin{figure}[h]
    \centering
    \vspace{-0.2cm}
    \includegraphics[width=0.48\textwidth]{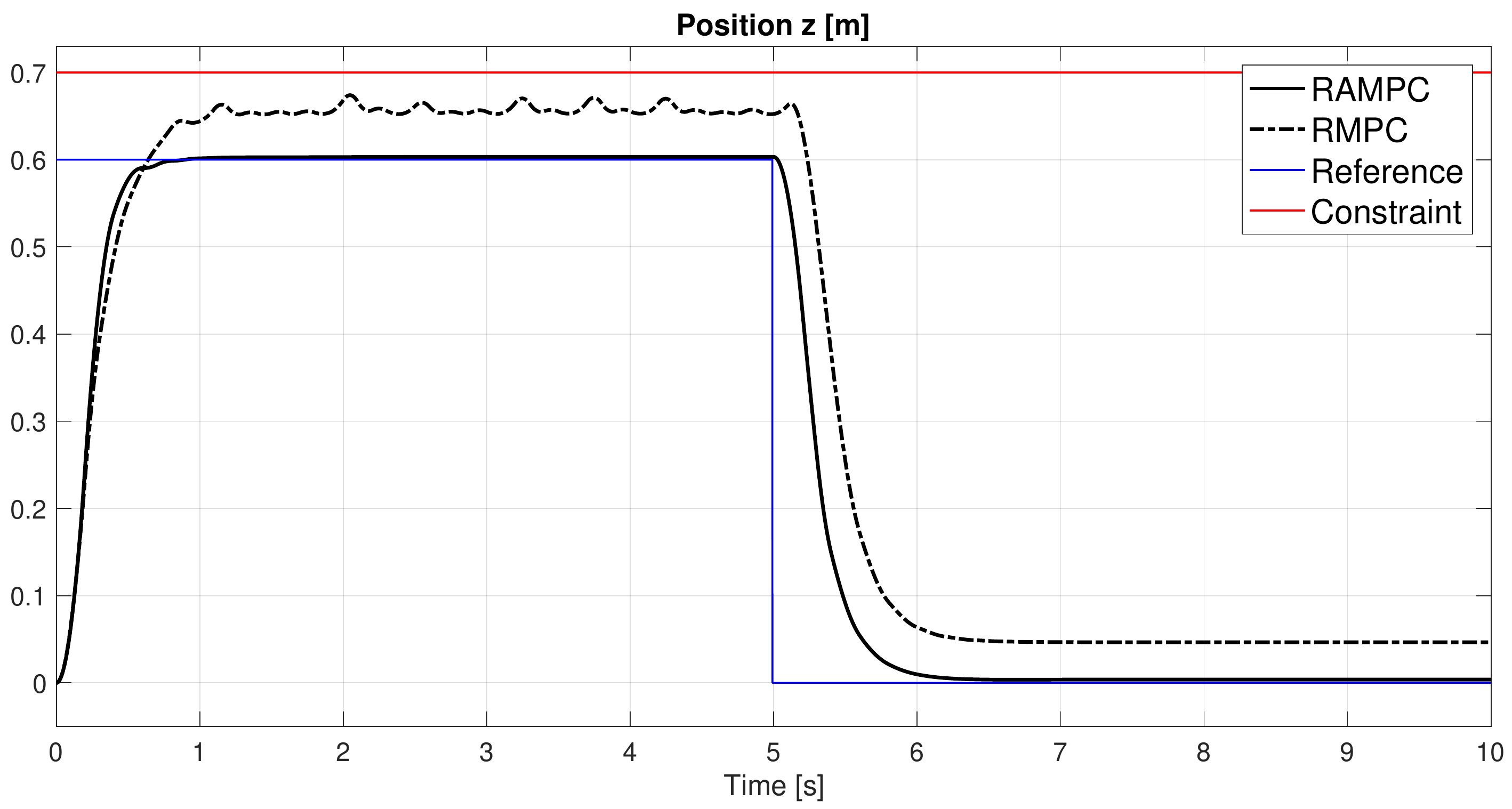}
    \vspace{-0.6cm}
    \caption{Comparison of robust MPC and RAMPC applied to a simulated quadrotor with an uncertain mass and wind as a disturbance with the altitude reference, constraint and position for robust MPC and RAMPC.}
\end{figure}

\subsection{Decoupled Quadrotor Dynamics}\label{sec:decoupleddynamics}
The linearisation in Section \ref{sec:quaddyn} represents one possible means of controlling a quadrotor via direct rotor control, i.e. the desired thrust for each rotor is computed individually and applied directly. This control mode proved to be difficult to implement, as discussed in Section \ref{sec:experiments}. Another possible mode of control decouples the quadrotor system into an $x$-position system, a $y$-position system, a $z$-position system and a yaw system which are controlled in an outer loop. The inputs in the decoupled system are the total thrust force deviation $\Delta f_{total}$ and $\Delta \omega_{x,\textup{ref}}$, $\Delta \omega_{y,\textup{ref}}$ and $\Delta \omega_{z,\textup{ref}}$ are the desired body rates about the $x$,$y$,$z$ body axes, respectively. 
The decoupled system thus needs a different controller for each subsystem. Since the scenarios considered in our simulation studies affect the $z$-direction the most, a RAMPC controller was chosen for $z$-position control while LQR controllers were used for the yaw and $x$- and $y$-position. This means that for this architecture, the subsystem controlled by RAMPC has 2 states and 1 control input, as given by
\begin{equation}
\begin{bmatrix} \Delta\dot{p}_z \\ \Delta \ddot{p}_z \end{bmatrix} = \begin{bmatrix}  0& 1\\ 0&0 \end{bmatrix} \begin{bmatrix}\Delta p_z \\ \Delta \dot{p}_z \end{bmatrix}+\begin{bmatrix} 0\\ \frac{1}{m}\end{bmatrix}\Delta f_{total}. \label{eq:decoupling}
\end{equation}
The desired body rates for the $x$-,$y$- and yaw-control are controlled in a linearised inner loop.
This inner control loop is controlled at a faster sampling rate by using a PID controller. 

\subsection{Unknown Mass Scenario with Decoupled Control}\label{sec:unknownmassdecoupled}
Using the decoupled control architecture, we first study the package delivery scenario described as in Section \ref{sec:unknownmass}. RAMPC is applied to the altitude control of a quadrotor, where the mass is unknown and its inverse lies in $\Theta_0=[\frac{1}{0.037},\frac{1}{0.027}]\textup{kg}^{-1}$ with $\theta^*=\frac{1}{0.028}$. The altitude is constrained to be within $0.7$m of the origin and the results of this simulation can be seen in Figure 3 as the dashed line, where the reference is tracked with no steady-state error.


\subsection{Power Delivery Failure Scenario} \label{sec:pdp}
The final scenario considered consists of a sudden power delivery failure for the decoupled altitude control. The rotor efficiency of all rotors $\gamma\in[0.7, 1]$ can drop at any given moment within these bounds and results in the dynamics $x_{k+1}=Ax_k+B\left(\frac{1}{m}\right)\gamma u_k.$
As the decoupled altitude control \eqref{eq:decoupling} is used, the uncertain parameter is
$\theta=\frac{\gamma}{m}$,
with $\Theta_0=[\frac{0.7}{0.037},\frac{1}{0.027}]$. In order to guarantee robustness for this failure at any time step, the lower bound $\theta_{k,\text{min}}$ of $\Theta_k$ is dilated at every time step after the parameter update according to $\theta_{k,\text{min}}=\min\{0.7\theta_{k,\text{min}},\theta_{0,\text{min}}\}$, where $\theta_{0,\text{min}}=\frac{0.7}{0.037}$. The result of such a failure is shown with a dashed line in Figure 4, where the failure occurs at 2 seconds. The applied RAMPC scheme manages to keep the quadrotor safe during the failure and with only a small steady-state error. This small steady-state error persists for the altitude due to the fact that the parameter set is dilated at every time step to ensure robustness against another failure.


\section{Experiments}
\label{sec:experiments}
\subsection{Experimental Configuration}
A Crazyflie quadrotor is used in the experiments. The mass of the Crazyflie is $27\textup{g}$ and it has a size of $92\times92\times29\textup{mm}$. It has a built in IMU consisting of 3 accelerometers and gyroscopes with up to a $2$kHz data rate. Through radio functionality, control commands are sent from a laptop, which computes the RAMPC solution, to the Crazyflie. 

In order to accurately track the position and rotation of the Crazyflie, a Vicon tracking system is used, which consists of 6 cameras placed in a room that track the motion of reflective surfaces attached to the Crazyflie. The data of the tracking system is collected using a separate computer and is sent to the laptop at a rate of $200$Hz. 

The MPC optimisation problem is solved using OSQP, see \cite{stellato2020osqp}, and takes less than $5\textup{ms}$. 
It was observed that the direct thrust control mode is difficult to implement on the Crazyflie, not only for RAMPC, but also with an LQR controller. Thus, the experiments were performed only with RAMPC in the decoupled altitude control mode in \eqref{eq:decoupling}. Although wind disturbances were not experimentally applied, the same disturbance bounds used in simulation were used to account for any linearisation error. Additionally, due to small measurement noise, the set-membership update with the dilated non-falsified parameter set \eqref{eq:SMnoise} is used.

\subsection{Unknown Mass Experiment}\label{sec:UME}
The unknown mass experiment is identical to the simulation configuration described in Section \ref{sec:unknownmassdecoupled} for the decoupled system. The RAMPC scheme is used for the altitude control of the quadrotor, while LQR controllers are used for the $x$ and $y$-position and yaw control. A PID controller stabilises the inner control loop \eqref{eq:decoupling} using the IMU of the Crazyflie. The quadrotor altitude is initially controlled using an LQR controller with an assumed mass of $37$g until RAMPC is activated at $t=3$s. RAMPC is able to identify the unknown mass of the quadrotor and follows the altitude reference as can be seen in Figure 3. However, compared to the simulation, the Crazyflie reaches the reference slower and a small steady-state error exists due to using the dilated non-falsified parameter set \eqref{eq:SMnoise}, which results in a slower size reduction of the set of estimates and the differences between the theoretical model and the Crazyflie. 
\begin{figure}[h]
    \centering
    \vspace{-0.3cm}
    \includegraphics[width=0.48\textwidth]{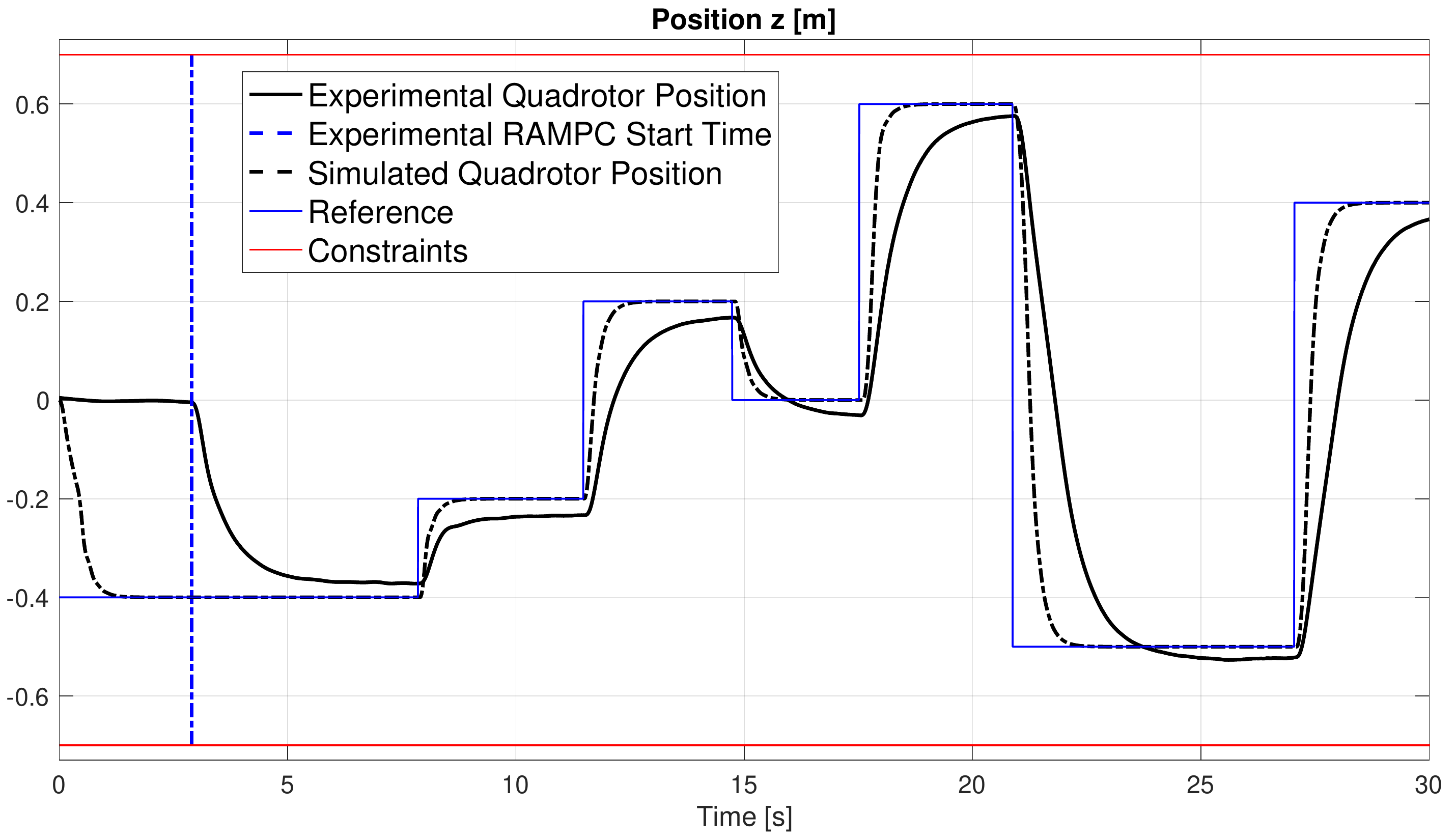}
    \vspace{-0.7cm}
    \caption{Comparison of a simulation and implementation of RAMPC for the unknown mass scenario. In simulation, RAMPC is applied to the altitude control of a quadrotor with an uncertain mass. In the implementation, RAMPC is applied to the same scenario and is activated after $t\approx3$s.}
\end{figure}
\begin{figure}[h]
    \centering
    \vspace{-0.6cm}
    \includegraphics[width=0.48\textwidth]{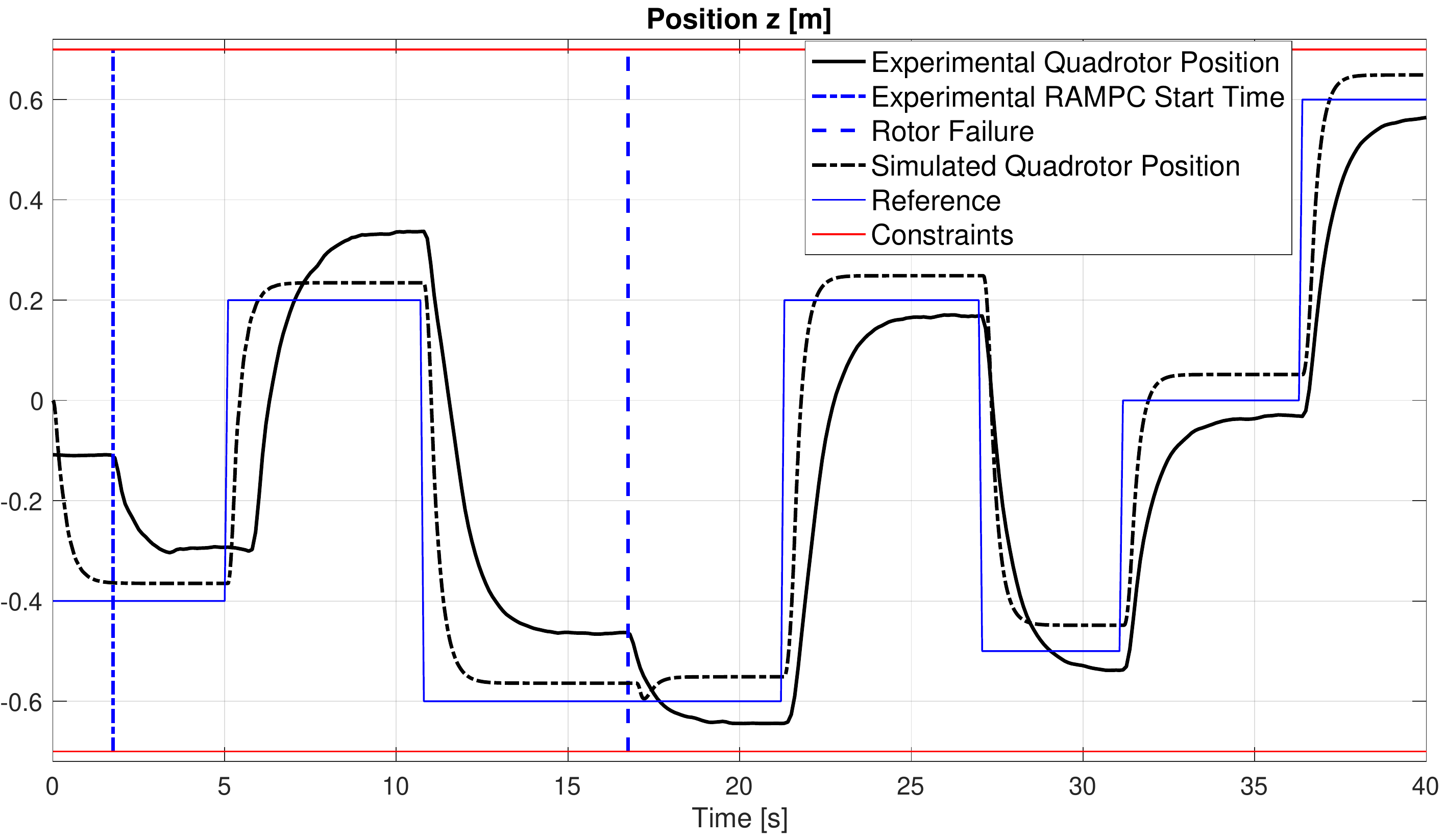}
    \vspace{-0.7cm}
    \caption{Comparison of a simulation and implementation of RAMPC for the power delivery failure scenario. In simulation, RAMPC is applied to the altitude control of a quadrotor with an uncertain mass and a power delivery failure after $t=17\textup{s}$ for all rotors, with $\gamma=0.7$. In the implementation, the same failure is considered with the RAMPC activation time after $t\approx2$s.}
\end{figure} 

\subsection{Power Delivery Failure Experiment}
This experiment is again the same as the configuration described in \ref{sec:pdp} for the decoupled system. The failure for this practical implementation, which can occur at any given time step, occurs at $t=17\textup{s}$. The occurrence of the failure is implemented by lowering the requested thrusts from the RAMPC scheme by $30\%$ on the quadrotor. As can be seen in Figure 4, the quadrotor recovers successfully from the failure and is able to track the given reference. Similarly to Section \ref{sec:UME}, small discrepancies exist in the tracking performance between simulation and experiment for the altitude control. 

\section{Conclusion}\label{sec:conslusion}
We used the Robust Adaptive Model Predictive Control scheme to run several experiments on a quadrotor. The existing RAMPC schemes were modified so that unknown steady-state inputs are considered and measurement noise is accounted for. The scenarios which were considered include an unknown mass experiment with wind as a disturbance and rotor failure of all rotors combined. For all the scenarios, it was shown through simulations that RAMPC managed to adapt to the uncertain parameters, as well as ensure state and input constraint satisfaction. Though the direct thrust control mode could not be experimentally implemented, it was shown that RAMPC can be applied to the resulting full state problem in simulations. RAMPC was then used to perform experiments for two proposed scenarios and the results were in agreement with those from the simulations. 

\addtolength{\textheight}{0cm}   





\end{document}